\documentclass[aps,epsf,twocolumn,eqsecnum]{revtex4}
\usepackage[dvips]{graphicx}
\usepackage{amsmath}
\skip\footins 15.25pt plus 4pt minus 2pt
\def\footnoterule{\kern-5.25pt\hrule width.5in\kern3.6pt}

\renewcommand{\mathrm}[1]{{\rm #1}}

\begin{document}
\draft


\title{\large\bf Hopping Models of  Charge Transfer in a Complex Environment:
New Class of Coupled Memory Continous-Time Random Walks}

\vskip 0.5cm

\vspace{24pt}



\author{ 	
Ewa {\sc Gudowska-Nowak}\ ${}^{1}$,
	Kinga {\sc Bochenek}\ ${}^{1}$,\\
         Agnieszka {\sc Jurlewicz}\ ${}^{2}$
	and Karina {\sc Weron}\ ${}^{3}$}

\vspace{8pt}



\address{
${ }^{1}$  {\sl Marian Smoluchowski Institute of  Physics 
Jagellonian University,
 ul. Reymonta 4,
30-059 Krak\'ow, Poland;}\\
${ }^{2}$ {\sl Hugo Steinhaus Center for Stochastic Methods and Institute of
Mathematics, Wroclaw University of Technology, 50-370 Wroclaw, Poland.}\\
${ }^{3}$  {\sl Institute of  Physics, Wroclaw University of Technology, 
	ul. Wybrze\.ze Wyspia\'nskiego 27, 50-370 Wroclaw, Poland.}}


\date{\today}

\vspace{6pt}

\begin{abstract}



Charge transport processes in disordered complex media are accompanied 
by anomalously slow relaxation for which usually a broad distribution of
relaxation times is adopted. To account for those properties of the environment,
a standard kinetic approach in description of the system is addressed either  in
the framework of continuous-time random walks (CTRW) or fractional diffusion.
In this paper the power of the CTRW approach is illustrated by use of the
probabilistic formalism and limit
theorems that allow to predict the limiting distributions of the paths traversed 
by charges and to derive effective relaxation properties of the entire system of
interest.
Application of the method is discussed for non-exponential electron-transfer
processes controlled by dynamics of the surrounding medium.  
\\ \\


\end{abstract}
\maketitle 
\newcommand{\gm}{\gamma}
\newcommand{\ee}{\epsilon}
\renewcommand{\th}{\theta}
\newcommand{\Sg}{\Sigma}
\newcommand{\dl}{\delta}
\newcommand{\SSg}{\tilde{\Sigma}}
\newcommand{\eq}{\begin{equation}}
\newcommand{\eqx}{\end{equation}}
\newcommand{\eqn}{\begin{eqnarray}}
\newcommand{\eqnx}{\end{eqnarray}}
\newcommand{\ben}{\begin{eqnarray}}
\newcommand{\een}{\end{eqnarray}}
\newcommand{\f}[2]{\frac{#1}{#2}}
\newcommand{\ra}{\rangle}
\newcommand{\la}{\langle}
\newcommand{\bra}[1]{\la #1|}
\newcommand{\ket}[1]{| #1\ra}
\newcommand{\GG}{{\cal G}}
\renewcommand{\AA}{{\cal A}}
\newcommand{\GR}{G(\ee)}
\newcommand{\MM}{{\cal M}}
\newcommand{\BB}{{\cal B}}
\newcommand{\ZZ}{{\cal Z}}
\newcommand{\DD}{{\cal D}}
\newcommand{\HH}{{\cal H}}
\newcommand{\RR}{{\cal R}}
\newcommand{\D}{\displaystyle}
\newcommand{\T}{\textstyle}
\newcommand{\St}{\scriptstyle}
\newcommand{\SSt}{\scriptscriptstyle}
\newcommand{\asarrow}{\raisebox{-1ex}[1ex][1ex]{\mbox{
\shortstack{${\St a.s.}$\\$\longrightarrow$\\[-0.3ex] 
${\SSt s\to\infty}$}}}}
\newcommand{\asarrown}{\raisebox{-1ex}[1ex][1ex]{\mbox{
\shortstack{${\St a.s.}$\\$\longrightarrow$\\[-0.3ex] 
${\SSt \delta\tau\to 0}$}}}}
\newcommand{\asarrownn}{\raisebox{-1ex}[1ex][1ex]{\mbox{
\shortstack{${\St a.s.}$\\$\longrightarrow$\\[-0.3ex] 
$^{}$}}}}
\newcommand{\darrow}{\raisebox{-1.2ex}[1ex][1ex]{\mbox{
\shortstack{${\St d}$\\$\longrightarrow$\\ 
${\SSt s\to\infty}$}}}}
\newcommand{\darrown}{\raisebox{-1.2ex}[1ex][1ex]{\mbox{
\shortstack{${\St d}$\\$\longrightarrow$\\ 
${\SSt \delta\tau\to 0}$}}}}
\newcommand{\darrownn}{\raisebox{-1.2ex}[1ex][1ex]{\mbox{
\shortstack{${\St d}$\\$\longrightarrow$\\ 
$^{}$}}}}
\def\del {\partial}
\def\stone {\mathop{\leftarrow}\limits}
\def\sttwo {\mathop{\rightarrow}\limits}
\def\koto {\sttwo^{\nu}}
\def\kuto {\stone_{\nu}}
\newcommand{\arr}[4]{
\left(\begin{array}{cc}
#1&#2\\
#3&#4
\end{array}\right)
}
\newcommand{\arrd}[3]{
\left(\begin{array}{ccc}
#1&0&0\\
0&#2&0\\
0&0&#3
\end{array}\right)
}
\newcommand{\tr}{\mbox{\rm tr}\,}
\newcommand{\One}{\mbox{\bf 1}}
\newcommand{\pauli}{\sg_2}
\newcommand{\cor}[1]{<{#1}>}
\newcommand{\cf}{{\it cf.}}
\newcommand{\ie}{{\it i.e.}}
\newcommand{\br}[1]{\overline{#1}}
\newcommand{\phib}{\br{\phi}}
\newcommand{\psib}{\br{\psi}}
\newcommand{\zb}{\br{z}}
\newcommand{\qb}{\br{q}}
\newcommand{\lm}{\lambda}
\newcommand{\ksi}{\xi}
\newcommand{\Gb}{\br{G}}
\newcommand{\Vb}{\br{V}}
\newcommand{\Gm}{G_{q\br{q}}}
\newcommand{\Vm}{V_{q\br{q}}}
\newcommand{\ggd}[2]{\GG_{#1}\otimes\GG^T_{#2}\Gamma}

\section{Introduction}

\noindent 
The stochastic formulation of transport phenomena in terms of a random walk
process, as well as the description {\it via} the deterministic diffusion
equation are two fundamental concepts in the theory of diffusion in complex
systems. The best known examples are charge transport in amorphous
semiconductors, rebinding kinetics in proteins, polarization fluctuations in
inhomogeneous solvents, diffusion of contaminants in
complex geological formations and diffusion of pollutants in large ecosystems.
In all realms mentioned above, the complex structures, characterized by a large
diversity of elementary units and strong interaction between them, exhibit a
nonpredictable or anomalous temporal evolution. The possibility of the dual
description  of the anomalous dynamical properties of such systems, based either
on the random motion or on the differential equations for the probability
density functions, has been considered in literature since the late 60s
and gave rise to an extensive list of developed models
\cite{MONTROLL,FRIEDRICH,HILFER}. \\
In this paper we demonstrate the power of the mathematical tools underlying the
concept of a continuous-time random walk (CTRW) by showing how the tool can be
generalized to handle complicated situations such as diffusion-reaction schemes
in complex system. The notion of the CTRW, a walk with a waiting time
distribution governing the time interval between subsequent jumps of a random
walker, has been introduced by Montroll and Weiss \cite{MONTROLL}.
The distribution of waiting times may stem from possible obstacles  and traps
that delay the particle's motion  and in consequence,  introduce the memory
effect into the kinetics. Especially fascinating in this approach was the idea
of an infinite mean time between the jumps as in such a case a characteristic time
scale of the process looses its common  sense. This novel concept has been 
used by
Montroll and Scher \cite{SCHER} to give a first explanation of experiments
measuring transient electrical current in amorphous semiconductors.
Since then the CTRW formalism has been successfully applied to describe fully
developed turbulence, transport in fractal media, intermittent chaotic systems
and relaxation phenomena. The common 
feature of the above mentioned  applications is that they
exhibit anomalous diffusion manifested by a non-Gaussian asymptotic distribution
(propagator, diffusion front) of a distance reached at large times.\\
At the level of the CTRW modeling, the diverging mean waiting time 
leads to a subdiffusive motion  with the mean square displacement
growing as $<r^2(t)>\propto t^{\alpha}$ with $0 <\alpha <1$. When applied to the
theory of Brownian motion, the CTRW scenario leads to the fractional diffusion 
equation \cite{METZLER,SOKOLOV} that can be treated on an equal footing with the framework
used for systems with normal diffusion.\\
Usually, in applications of the CTRW ideology, the analysis of the asymptotic
distribution is presented within the approach that is based on a formal
expression for the Fourier-Laplace transform of the propagator, or otherwise,
use of the fractional calculus is required \cite{HILFER} as a legitimate tool.
 Here, we present an approach to a random walk analysis which is based directly
on the definition of the cumulative stochastic process. Our aim is to show that
despite of the extensive studies on CTRWs and their long history in physics,
the powerful tool of the limit theorems \cite{ZOLOTAREV} hidden behind the
derivation of limiting distributions, has not been fully explored yet.
 We emphasize the possibilities of applications of that scenario in stochastic
modeling of physical systems, in particular, in description of  the charge
transport in disordered materials.

 A starting point in the CTRW analysis is the definition of a total path $R(t)$
of a particle traversed  up to the time $t$ in accumulating number $L(t)$ of 
jumps of a length $R_j$.
The number of jumps exerted in (generally random) time $t$ can be defined
either directly by assuming a specific counting process $L(t)$ (with {\it e.g.}
Poisson, negative binomial, geometric, etc. count distribution) or indirectly -
by assuming the distribution of waiting times $T_j$ between the jumps. 
In both approaches, under certain assumptions concerning the distribution of
jumps $R_j$ and the distribution of waiting times $T_j$ (or number of jumps
$L(t)$), the asymptotic distribution of the total path $R(t)$ reached up to time
$t$ can be obtained by applying limit theorems of the probability theory.
In contrast to the more popular Tauberian analysis of the Fourier-Laplace transform
of $R(t)$, such an approach precisely identifies classes of possible limiting
distributions and offers an easy-to-follow scheme of generating various
limiting results. \\
The dual description of the random walk and the relationship between
the results obtained in both cases are exemplified in this work by discussing
the biologically relevant charge transport processes in disordered media.

 The paper is organized as follows: We begin in Section II with a brief
 discussion of models of
 non-exponential dielectric relaxation and their relation to solvent (medium) dynamics 
 influencing
 the rates of the long-range electron transfer. Further, as a generalization of
 the McConnell formula we incorporate medium  fluctuations in the expression for
 the electronic transfer matrix. Its form is analyzed in terms of an exponential
 of a sum of independent and identically distributed ({\it {\it i.i.d}}) random variables
  with a random number $L$ of virtual jumps between the donor
  and acceptor sites. By assuming the deviations from equilibrium of the atomic
  coordinates of a given  pathway to be  random contributions to the sum, we are able
   to investigate asymptotic forms of the tunneling matrix elements. Section III
   poses the problem in terms of a standard CTRW scenario which is generalized
   (Sections IV and V)  for random walks subordinated to a compound step-counting
   process. Main results and conclusions of the analysis are presented in 
   Sections VI and VII.
   
\section{Charge transport in a complex environment}
Charge transport processes  
determine a variety of phenomena in
physics, chemistry and biology. The study of the phenomenon  
has gradually developed together with
general progress in theoretical physics and in fast high resolution 
spectroscopy, so that contemporary research deals
nowadays with a broad class of systems, materials and environmental conditions.
Of particular interest are the processes taking place in disordered materials \cite{RICHERT},
such as amorphous semiconductors, randomly arranged molecular wires, glasses or biological
proteins where the charge transfer processes form the elementary steps in
energy transport and  production in almost all living  cells.
In all those cases, the actual transport process is coupled to local
polarization  fluctuations of the environment. For the situations that the
relaxation of the polarization fluctuations of the surrounding medium has a
simple ``close-to-equilibrium'' exponentially decaying form, the main energetic
contributions to the charge transfer process come from the reorganization
energy of the medium \cite{ULSTRUP,MARCUS}. 
In contrast, many observed charge transport processes, like electron transfer
(ET) in complex solvents \cite{COLE,METZLER,DAV,HAASE,WIL,HAV,JONSCHER,RICHERT} and 
proteins \cite{ONUCHIC,PANDE,BALABIN}, or gating kinetics of biological
channels \cite{NADLER}, exhibit non-exponential kinetics resulting from the complex response
to the interfering medium. A classical example are higher alcohols, for which 
the frequency dependent dielectric permittivity takes on a Cole-Davidson (CD)
\cite{DAV} form:
\ben
\phi^{*}_{CD}(\omega)=\frac{\epsilon^*(\omega)-\epsilon_{\infty}}{\epsilon_0-\epsilon_{\infty}}=
\frac{1}{(1+i\omega\tau_p)^{\gamma}}=\nonumber \\
=\int^{\infty}_0 e^{-i\omega t}\left [-\frac{d}{dt} \phi(t/\tau_p)\right]dt
\label{perm}
\een
with $0<\gamma<1$; $\phi(t)$ standing for the correlation function of polarization 
fluctuations and $\tau_p$ indicating a reciprocal proportional to the peak
frequency of the dielectric loss.
In the electron transfer (ET) theory, the time-correlation function $\phi(t)$ is related
\cite{MARCUS,ULSTRUP,HYNES} to the Coulombic potential energy difference for a
given configuration of all solvent (intervening medium)  molecules in the
states of reactants and products:
\ben
\phi(t)=\left<(\delta\Delta E)^2\right>^{-1} \left <\delta\Delta E\delta\Delta
E(t)\right>
\een
with $\Delta E(t)$  identified with a complex dynamic ``reaction coordinate''
describing the transfer.
In a convenient dipole-approximation for medium molecules, the potential energy
difference $\Delta E$ would be given by \cite{HYNES}
\ben
\Delta E=-\int d{\bf r} {\bf P}({\bf r}) [E_P({\bf r-r_P})-E_R({\bf r-r_R})]
\een
where ${\bf P}({\bf r})$ stands for the medium orientational polarization at
position ${\bf r}$. For a solvent in which the dipoles of the
dielectric medium relax with a single relaxation time $\tau_p$, the complex
dielectric permittivity Eq.(\ref{perm}) is given by the Debye (D) function
\ben
\label{debye}
\phi^{*}_{D}=\frac{\epsilon^*(\omega)-\epsilon_{\infty}}{\epsilon_0-\epsilon_{\infty}}= \frac{1}{(1+i\omega\tau_p)} 
\een
with $\phi(t)$ expressed in terms of a single exponential function with a decay
time $\tau_p$.
Other, equally likely fitted expressions \cite{JONSCHER,HAASE} exploited in dielectric
spectroscopy of polymers and disordered solids estimate relaxation of $\phi(t)$
by use of the Cole-Cole (CC) \cite{COLE} formula
\ben
\label{cole}
\phi^{*}_{CC}=\frac{\epsilon^*(\omega)-\epsilon_{\infty}}{\epsilon_0-\epsilon_{\infty}}=
\frac{1}{1+(i\omega\tau_p)^{\alpha}}
\een
or the Havriliak-Negami (HN) \cite{HAV} function:
\ben
\label{havriliak}
\phi^{*}_{HN}=\frac{\epsilon(\omega)-\epsilon_{\infty}}{\epsilon_0-\epsilon_{\infty}}=
\frac{1}{(1+(i\omega\tau_p)^{\alpha})^{\gamma}}
\een
where $0<\alpha<1$ and $0<\gamma<1$ are parameters determining the characteristics of the
dielectric relaxation with $\alpha$ representing the width 
and $\gamma$ the skewness of the
distribution of relaxation times \cite{HAV}. 
Although the generic physical reasons for anomalous
relaxation in complex systems are still under debate, both - static models
based on the inhomogeneity of the medium, - as well as the dynamic models,
describing  complex local dynamical processes have been successfully employed to describe 
relaxation behavior of fluctuations in such systems. In particular,
the studies on the effect of protein dynamics on biological ET
\cite{ULSTRUP,STUCH,ONUCHIC,NEWTON,KESTNER} have demonstrated sensitivity of the long distance
tunneling mediated by the protein matrix on atomic configurations of the 
surroundings and pointed out possibility of an electron of emitting or
 absorbing phonons from the medium that would effectively result in an
inelastic ET processes.
In numerous chemical and biological
examples of the ET reaction~\cite{NEWTON,ULSTRUP,KESTNER}, a single electron is tunneling in 
an inhomogeneous medium over large distances of several angstroms. The
intervening medium can be either a protein backbone or a sequence of
cofactors embedded in a protein matrix.  Due to a large separation
between the donor and acceptor, direct electronic coupling between the
chromophores is negligible, rendering thus the question on the effect of
medium on enhancement of the electronic coupling~\cite{ONUCHIC}. 
A possible realization  of the long-distance ET process is a transfer
mediated through the medium which acts as a bridge providing virtual states for 
the tunneling
electron~\cite{CONNEL}.
Within the nonadiabatic-reaction scenario corresponding to a weak electronic
coupling $T_{DA}$ between the state of reactants $D$ and products $A$, the 
expression for the rate reads
\ben
k_{ET}=\frac{2\pi}{\hbar}T^2_{DA}(FC) 
\een
where (FC) is the Franck-Condon nuclear factor associated with the nuclear modes activation barrier. In a conventional theory the Condon approximation is assumed, {\it i.e.} the electronic coupling $T_{DA}$ is viewed as independent of the coordinates of the medium.
To account for thermal fluctuations of the bridge or random intervening medium,
the electronic coupling has to be a function of the modes of the medium.
The simplest expression that can be proposed in such a case is the Mc Connell
formula~\cite{CONNEL,STUCH} which 
for a  case of a linear bridge consisting of $L$ orbitals leads
to the tunneling matrix $T_{DA}$ 
\ben
T_{DA}\approx \prod^L_j\frac{\beta_{j,j+1}}{\ee-\ee_j}
\een
with $\ee-\ee_j$ being the energy difference between the tunneling energy and 
the energy of the 
bridging orbital $j$, $L$ standing for the number of virtual jumps performed along the path
 and $\beta_{ij}$ denoting  couplings between directly
overlapping atomic orbitals of neighbouring atoms within the bridge. 
The above formula constitutes the essential part of the ET pathways models \cite{ONUCHIC,STUCH}
in proteins, where the calculation of the effective electronic coupling is based on a general
assumption that the electron wave function decay is softer for propagating through a chemical bond
than through space jump. Since the coupling coefficients $\beta$
are exponentially decaying function of the distance between subsequent medium centers (atoms), the effective 
tunneling matrix can be recast in the form
\ben
T_{DA}(r)=T_{DA}^0\prod^L_j \exp(-\alpha_j r_j)=\nonumber \\
=T_{DA}^0\exp\left(-\sum^L_j
\alpha_j r_j\right)\nonumber \\
=T_{DA}^0\exp \left (-\sum^L_j R_j\right)
\een  
where $r_j$ are fluctuations of the atomic coordinates of the bridge, 
$\alpha_j$ are constants characterizing strength of the coupling to a 
particular bridge mode $j$ and $T_{DA}^0$ corresponds to the average, 
equilibrium tunneling matrix.
Such a representation of the effective 
tunneling matrix  allows  to use the notion of the continuous 
time random walk (CTRW) as a very convenient mathematical tool to analyze 
the decay with time of the donor state occupation density. In fact, 
the latter is
commonly described by systems of phenomenological balance equations
\ben
\frac{d}{dt} \left( \begin{array}{c}  P_1(t) \\ P_{2}(t) \end{array} \right) = 
-\left( \begin{array}{cc}  k^+ & -k^-\\-k^+ & k^- \end{array} \right)\left( \begin{array}{c} 
P_1(t) \\ P_{2}(t) \end{array} \right)
\label{ma}
\een
which relate the decay of the donor (acceptor) populations to the
state-relaxation 
rate constants $k^{+,-}$. Note, that corresponding populations $P_{1,2}(t)$
in any of the electronic states (reactants or products) are dynamical 
quantities usually measured in the electron transfer kinetic
experiment and are obtained by integrating the polarization energy dependent
populations $\rho(E,t)$ over configuration variable $E(t)$:
\ben
P_i(t)=\int^{+\infty}_{-\infty}dE \rho(E,t)
\een
 In a standard ET theory approach \cite{HYNES,STUCH,NEWTON}
after assuming a disentanglement of reactive tunneling from the dynamics of
diffusion, the elements of the evolution matrix Eq.(\ref{ma}) have the form of
\ben
k^+=\frac{k^+_{NA}}{1+ k^+_{NA}/k^+_D+k^-_{NA}/k^-_D}
\een
where $k_{NA}$ describes the crossing (nonadiabatic) kinetics and $k_D$ is the rate
constant
 of the diffusion in the reactants' (products') basins. However, in a more
general situations, where the matrix entries in Eq.(\ref{ma}) are represented by
time dependent functions, the redistribution of populations and consequently, the
relaxation of electron-donoring (accepting) states may follow a
non-exponential law. Accordingly, the frequency characteristics of
dielectric susceptibility 
$\chi(\omega)$
connected to the temporal relaxation function of the induced state-polarization
${\bf P}(t)=\epsilon_0\chi(\omega) {\bf E}(t)=\epsilon_0(\epsilon^*-1) {\bf
E}(t)$ where ${\bf E}(t)={\bf E}_0e^{-i\omega t}$ and the
functional
character of the dielectric permittivity $\epsilon^* (\omega)$
may be inferred from the analysis of relaxation of state populations in a frequency domain
\ben
\chi(\omega)=\int^{\infty}_0e^{-i\omega t} d(-P_{1,2}(t))
\een 																												
In the forthcoming sections we present a dynamic framework which, within the
 CTRW cenario, leads to the empirically observed non-exponential relaxation 
 dynamics.
\section{Pathway analysis of ET reactions}
As discussed above, with the matrix elements of a particular path
$T_{DA}$, the nature of disorder may be analyzed in terms of fluctuations in
couplings or, alternatively,  in contributions $R_j$ to the total  distance $R$ 
traversed by a charge:
\ben
R=\sum^{L(\cal{R})}_{j=0}R_j\ge\cal{R}
\label{distance}
\een 
Here $L(\cal{R})$ stands for a random counting process describing a (random) number of
forward steps exerted by a particle before reaching a distance $\cal{R}$. Note,
that such a formulation is identical with the assumption of a one-dimensional, biased (directed)
random
walk performed in an amorphous medium under the influence of a strong external
field. The time $T$ that particle needs to reach a fixed distance $\cal{R}$ is
given by $T=\sum^{L(\cal{R})}_{i=0}T_i$ with $T_i$ being a time spent by a
hopping  charge at the location $\sum_j^{i-1} R_j$. By means of a conditional
probability, the probability density $p(T)_{\cal{R}}$ for the distribution of
times $T$ to reach a distance $\cal{R}$ reads
\ben
p(T)_{{\cal{R}}}=\nonumber \\
=\sum^{\infty}_{j=1} p_1(L({\cal{R}})=j)
p_2\left (\sum^{L({\cal{R}})}_{i=0}T_i=T|L({\cal{R}})=j \right )
\label{prob}
\een
where $p_2$ stands for the probability distribution that the elapsed time is $T$
provided exactly $j$ steps have been performed to reach distance $\cal{R}$ and
 $p_1$ is
probability to make $j$ steps over the distance $\cal{R}$.
Here we assume, that the length of a given jump, as well as the waiting times
elapsing between two successive jumps are drawn as independent random variables
with densities
\ben
\rho(r)=\beta e^{-\beta r},\;\;r>0,
\een 
and
\ben
\sigma(t)=L_{1/2}(t;1,1,0)=\frac{1}{\sqrt{2\pi}}t^{-3/2}e^{-1/2t},\;\;t>0,
\label{czas}
\een
i.e. the model describes a charge moving only in one direction with a Poisson
number of jumps and the L{\'e}vy-Smirnov distribution $L_{1/2}(t;1,1,0)$ of  waiting times. 
The probability density function $p_2$ in Eq.(\ref{prob}) is
then given by a convolution of independent probability densities (\ref{czas}) 
\ben
p_2(T)=L_{1/2}(T;j^2,1,0)
\een
and leads to a marginal probability density 
\ben
p(T)_{{\cal{R}}}=e^{-\beta{\cal{R}}}\frac{T^{-3/2}}{\sqrt{2\pi}}\sum^{\infty}_{j=0}\frac{(\beta{\cal{R}})^j}{j!}j
e^{-\frac{j^2}{2T}}
\een

\begin{figure}[h]
\includegraphics[height=4.5cm, width=8.5cm]{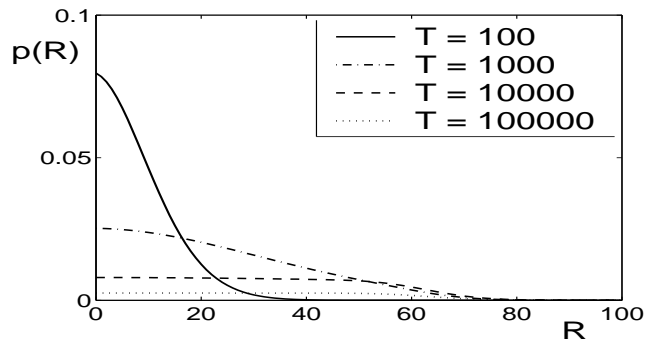}
\caption{Density distribution function $p(R)_{{\cal{T}}}$ of distances traversed
up to the time ${\cal{T}}$ 
in a one-dimensional random with 
 $\tau(t)$ given by a
L\'evy-Smirnov distribution Eq.(\ref{czas})}. 
\label{amp}
\end{figure}
\begin{figure}[ht]
\includegraphics[height=4.5cm, width=8.5cm]{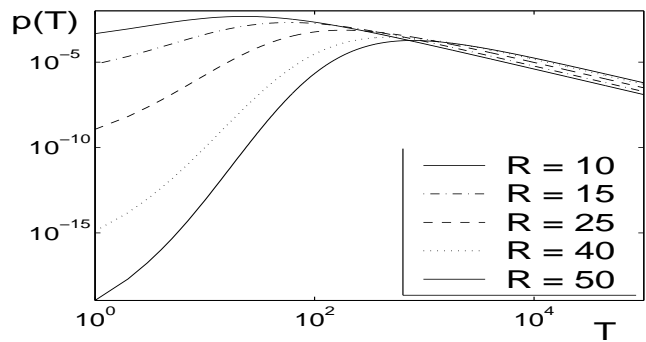}
\caption{Density distribution function of times $T$ needed to reach a distance
${\cal{R}}$ in a one-dimensional random walk with a L\'evy-Smirnov distribution
 Eq.(\ref{czas}) of waiting times for a jump.}
\label{pamp}
\end{figure}
\begin{figure}[ht]
\includegraphics[height=8.5cm, width=8.5cm]{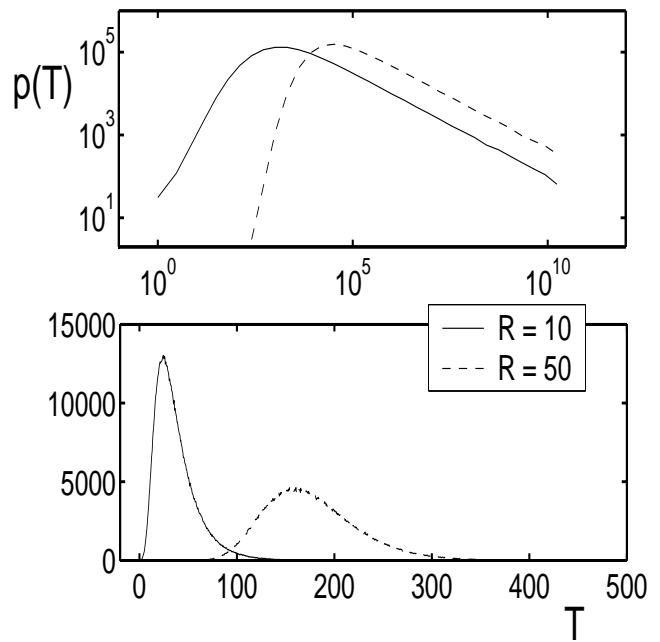}
\caption{Density distribution function of times $T$ needed to reach a distance
${\cal{R}}$ in a three-dimensional CTRW with a L\'evy-Smirnov Eq.(\ref{czas})
distribution (upper panel) and
a Weibull distribution Eq.(\ref{czas1} (lower panel)
of waiting times for a jump.}
\label{pamp1}
\end{figure}

Quite similarly, one can calculate $p(R)_{{\cal{T}}}$ with the condition
$T=\sum^{L({\cal{T}})+1}_{j=1}T_j\ge{\cal{T}}$ and $p_2$ given by a gamma
density distribution resulting from a sum of $j$ independent and exponentially 
distributed random variables:
\ben
p(R)_{{\cal{T}}}=\nonumber \\
=\sum_{j=1}^{\infty}
p_1(L({\cal{T}})=j)p_2\left (\sum^{L({\cal{T}})}_{i=0}r_i=R|L({\cal{T}})=j
\right )=\nonumber \\
=\sum_{n=0}^{\infty}\frac{(\beta {\cal{R}})^n}{n!}e^{-\beta
{\cal{R}}} \frac{2}{\sqrt{2\pi T}}\int^{n+1}_ne^{-\frac{y^2}{2T}}dy\nonumber \\
\een
\label{prob1}
Thus the sketched problem describes a 1-dim diffusion among traps with a broad
(asymptotically, heavy-tailed) distribution $\sigma(t)$ of trapping
times, see Figures~1 and~2. This
type of the ``annealed'' CTRW has been extensively studied in literature
\cite{MONTROLL,WEISSMAN,KLAFTER,HAASE,RICHERT}. A 3-dim analogue with a broken unidirectionality of
the transfer corresponds to a weak external field
approximation.  Figure~\ref{pamp1} displays results of a 3-dim CTRW
computer
simulation: direction of a jump has been generated by sampling spherical
coordinates $\theta$ and $\phi$ from uniform distributions defined on intervals
$(0,\frac{\pi}{2})$ and $(0, 2\pi)$, respectively. Quite arbitrarily, the
positive direction of the $z$-axis 
have been favoured by sampling the coordinates of a point $(\phi,\theta, -z)$
by using those for $(\phi,\theta, z)$ and switching from $z$ to
$-z$ with a probability 3/10. The upper panel of Figure~\ref{pamp1} refers to
the distribution of times $T$ required to reach the distance $R$ when the
distribution of waiting times for the subsequent jump was assumed in a form of
a skewed (heavy-tailed) L\'evy-Smirnov distribution describing higher
probability of long trapping. In contrast, the lower panel represents results of
simulations 
for a 3-dim CTRW with a preferential short mean time of waiting (a Weibull
distribution) for a particle release from the trap:
\ben
\sigma(t)=\alpha t^{\alpha-1}e^{-t^{\alpha}}
\label{czas1}
\een
where $0<\alpha<1$.
In a forthcoming section we investigate an extension of such an approach for
the family of complex CTRW  incorporating effects of clustering of carrier
trapping sites in a disordered material under the study.  

\section{CTRW subordinated to a compound counting process}
As briefly mentioned in the preceding sections, the  CTRW generalizes a simple random walk by implementing 
a random waiting time between jumps \cite{MONTROLL,SCHER}. This stochastic 
process is defined by the total distance $R(\tau)$
reached by the particle at time $\tau\geq 0$
if the movement is generated by a sequence  
$\{(R_{j},T_{j}), j=1,2,\ldots\}$ of jump parameters with 
$R_{j}$ specifying both the length
and the direction of the $j$th jump and $T_{j}>0$ denoting the waiting 
time for the next jump. 
We assume moreover that the sequence
$(R_{j},T_{j}), j=1,2,\ldots$ is formed by {\it {\it i.i.d}.}~random vectors.
The cumulative distance $R(\tau)$ may be (cf.~Eq.~(\ref{distance})) expressed as
\ben
R(\tau)=\sum\limits_{j=1}^{\nu (\tau)} R_{j}\nonumber \\
\een
where
\ben
\nu (\tau)=\min\{k: \sum\limits_{j=1}^{k} T_{j}>\tau\}
\label{1}
\een
The renewal process
\ben
\left \{ \sum_{j=1}^{k} T_{j}, k=1,2,\ldots \right \}
\een
represents the instants
of time at which subsequent jumps occur; the process 
$\{\nu (\tau), \tau\geq 0\}$ counts the jumps.  
Hence, the introduced CTRW $\{R(\tau), \tau\geq 0\}$ is the discrete-time 
random walk $\{ \sum\limits_{j=1}^{[\tau]} R_{j}, \tau\geq 0\}$ 
subordinated to the renewal counting  process 
$\{\nu (\tau), \tau\geq 0\}$.
It is  decoupled if random variables $T_{j}$ and $R_{j}$
are independent; otherwise it incorporates statistical dependence
between time and space steps. 
In general, the properties of $\{R(\tau), \tau\geq 0\}$ 
are unknown. Nevertheless, the behavior of the CTRW for long time, or,
equivalently, of the rescaled process $R(t /\delta \tau)/f(\delta \tau)$, 
(where 
$f(\delta\tau)$ is an appropriately chosen rescaling function) for 
the characteristic time scale $\delta\tau$ decreasing to 0,
can be determined quite well. 
Systematic studies of the limiting
total-distance distributions for the one-dimensional
CTRWs have pointed on few possible distributions \cite{WERON,KOT}.
Introducing a new class of coupled memory CTRWs with random-sum form of
time/space steps, 
much broader 
class of possible limiting distributions can be obtained.
Namely, let us consider the case when  
\ben
\label{xiii}
R_{j}=\sum\limits_{k=1}^{M_{j}}\Delta R_{jk}, \quad 
T_{j}=\sum\limits_{k=1}^{M_{j}}\Delta T_{jk},
\label{def}
\een
where sequences $\{\Delta R_{jk}, j,k=1,2,\ldots\}$ and 
$\{\Delta T_{jk}, j,k=1,2,\ldots\}$ of space/time random spans 
are independent and each of them
consists of {\it i.i.d}.~positive random variables, and 
$\{M_{j}, j=1,2,\ldots\}$ is a sequence of {\it i.i.d}.~positive 
integer-valued random variables. 
The nondegenerate distribution of $M_{j}$, the number of summands,
(i.e.~the case when random variable $M_{j}$ takes at least two 
different values with positive probabilities),
provides a stochastic dependence between time and space steps. 
As a consequence, the CTRW resulting from the space/time-step family 
Eq.(\ref{xiii}) is usually coupled. 
Moreover, it has an equivalent random-sum form since
the total distance $R(\tau)$ has the same distribution as 
the following random sum of distance increments
\ben
R(\tau)=\sum\limits_{k=1}^{L(\tau)} \Delta R_{1k}
\label{suma1}
\een
where the random index $L(\tau)$ has the form 
\ben
L(\tau)=\sum\limits_{j=1}^{N(K(\tau))} M_{j},
\label{suma2}
\een
with the random number of summands defined as 
\ben
N(K(\tau))= \min \{n: \sum\limits_{j=1}^{n}M_{j}>K(\tau)\}
\een
for
\ben
K(\tau)= \min \{n: \sum\limits_{k=1}^{n}\Delta T_{1k}>\tau\}. 
\een
Note, that the above definition of the CTRW process (cf. Eq.(\ref{suma1})) does not
assume explicitly any precise relation between steps $R_i$ and time lapses
$T_i$. Instead, possible coupling between those two random variables is
incorporated {\it via} the counting process Eq.(\ref{suma2}) summing 
contributions to the cumulative distance $R(\tau)$ traversed during the walk. 
This particular construction of the random sum Eq.({\ref{suma1}) allows
calculating statistical properties of their distribution by use of the
``nested'' moment-generating function $\Phi(s)=< s^{R(\tau)}>$ which 
in this case fulfills the equation
\ben
\Phi_R(s) =\Phi_L(\Phi_{\Delta R_{1k}}(s))=\Phi_N(\Phi_M(\Phi_{\Delta
R_{1k}}(s)))
\een
where the last equation follows by applying the technique of conditional
averages.
\section{Diffusion front. Limiting distributions}
Let us consider now the rescaled total distance 
\ben
{\widetilde R}_{\delta \tau}(t)=\D{R_{M}(t /\delta \tau)\over f(\delta
\tau)}
\label{reg}
\een
where $\delta\tau$ is  the characteristic time scale, and 
$f(\delta \tau)$ is a rescaling function chosen appropriately.
Applying limit theorems \cite{meerschaert} one can evaluate the limiting
position ${\widetilde R}(t)$ of the rescaled total distance reached as
$\delta\tau\to 0$. 
The characteristics of ${\widetilde R}(t)$ depends on
assumptions set on the distributions of the
variables $\Delta R_{jk}, \Delta T_{jk})$ and $M_{j}$.
Below, following the regularization scheme Eq.(\ref{reg}) along with a more
detailed presentation and proofs included in
ref.\cite{ajurl2003,meerschaert},
we discuss briefly some examples which might be of practical use
in modeling relaxation phenomena in disordered materials. We consider 
the case when $\Delta R_{jk}$ and $\Delta T_{jk}$ are independent and both
positive.

\begin{enumerate}
\item [(a)]
Let us first assume that both 
$\Delta R_{jk}$ and $\Delta T_{jk}$
have heavy-tailed distributions with 
$c=c_{1}$ and $c=c_{2}$, respectively, and the same $r=\alpha$. (We say 
that the distribution of a positive random variable $X$ 
has a {\it heavy tail} if for some $c>0$ and $0<r<1$
\ben
\lim\limits_{x\to\infty}\D{\Pr (X\!>\!x)\over (x/c)^{-r}} 
= 1.
\een
In consequence, the expected value $\langle X\rangle$ 
is infinite.) 
\begin{itemize}
\item If the distribution of the random number  $M_{j}$ 
has a heavy tail with some $c>0$ and $r=\gamma$,
then for any $t>0$
\ben
\label{4}
\D{R(t /\delta \tau)\over 1/\delta \tau}
\darrown {\widetilde R}(t)\;\stackrel{d}{=}\;
\D{t\over A}\;
\D{{\cal S}_{\alpha}^{'}\over {\cal S}_{\alpha}}
\left (\D{1\over {\cal B}_{\gamma}}\right )^{1/\alpha}.
\een
Here ''$\darrownn$'' reads ''tends in distribution'', and 
''$\stackrel{d}{=}$'' denotes the equal distributions.
\item  If the numbers $M_{j}$ have a finite expected value ($\langle
M_{j}\rangle <\infty$), then for any $t>0$
\ben
\label{5}
\D{R(t /\delta \tau)\over 1/\delta \tau}
\darrown {\widetilde R}(t)\;\stackrel{d}{=}\;
\D{t\over A}\;
\D{{\cal S}_{\alpha}^{'}\over {\cal S}_{\alpha}}.
\een
\end{itemize}
\item [(b)]
Assume that the expected values of both $\Delta R_{jk}$ and $\Delta T_{jk}$
are finite, and 
$\langle \Delta R_{jk}\rangle =c_{1}$, $\langle \Delta T_{jk} \rangle =c_{2}$.
\begin{itemize}
\item If the distribution of 
$M_{j}$ has a heavy tail with some $c>0$ and $r=\gamma$,
then for any $t>0$
\ben
\label{6}
\D{R(t /\delta \tau)\over 1/\delta \tau}
\darrown {\widetilde R}(t)\;\stackrel{d}{=}\;
\D{t\over A}\;
\D{1\over {\cal B}_{\gamma}}.
\een

\item  If $\langle M_{j}\rangle <\infty$, then for any $t>0$
\ben
\label{7}
\frac{R(t /\delta \tau)}{1/\delta \tau}
\asarrown {\widetilde R}(t)\nonumber \\
=\frac{1}{A}
\een
Here ''$\asarrownn$'' reads ''tends with probability 1''.
\end{itemize}
\end{enumerate}

The random variables ${\cal B}_{\gamma}$, ${\cal S}_{\alpha}$, and
${\cal S}_{\alpha}^{'}$ in (\ref{4}) - (\ref{7}) are as follows:\\
\begin{itemize}
\item ${\cal S}_{\alpha}$ and ${\cal S}_{\alpha}^{'}$
are identically distributed according to the completely asymmetric
$\alpha$-stable law such that   
$\langle e^{-k{\cal S}_{\alpha}}\rangle= e^{-k^{\alpha}}$;
\item ${\cal B}_{\gamma}$ is
distributed according to the generalized arcsine distribution with
parameter $\gamma$ (i.e., the beta distribution with parameters
$p=\gamma$ and $q=1-\gamma$) given by the density function
\ben
f_{\gamma}(x) = 
\begin{cases}
\frac{1}{\Gamma(\gamma)\Gamma(1-\gamma)}
x^{\gamma-1}(1{-}x)^{-\gamma}& \text{for}\ 0<x<1\\
0&\text{otherwise};
\end{cases}
\een

\item  for any $0<\alpha,\gamma < 1$ the 
random variables ${\cal B}_{\gamma}$, ${\cal S}_{\alpha}$, and
${\cal S}_{\alpha}^{'}$ are independent.
\end{itemize}
Armed with the above results, we are now in position to discuss
properties of CTRW paths generated under mentioned constraints with
the application of the formalism in the analysis of the relaxation responses
in disordered materials.

\section{Empirical and phenomenological relaxation responses}
Relaxation in amorphous solids and ET processes in disordered molecular media
represent nowadays
intensively investigated subjects both in experimental and theoretical
physics \cite{JONSCHER,METZLER,WERON,PANDE,SCHLICHTER,US}. 
In particular, a key probe of electron dynamics in disordered systems is the
time of flight experiment (TOF) for the drift mobility. In the experiment, the
thin film sample is located between two blocking contacts across which is
maintained a potential drop, and a laser flash is used to create carriers that
wander towards an appropriate electrode. During their drift through the sample, the
electrons and holes encounter a variety of traps that affect their motion.
The experiments show that in the disordered materials, the registered 
transient current follows an algebraic decay $I(t)\approx t^{-(\alpha +1)}$.
In contrast, for Gaussian transport processes, the charge carriers move at a
constant velocity and after a transient time, depending on the thickness of the
sample and the applied external field, they become absorbed. In consequence, for
normal transport processes typically observed in ordered materials, the current is given
by a step-like function, whereas in disordered media the current $I(t)$ adheres
to a universal (independent of
the applied field and sample thickness) scaling curve.
Similar conclusions are drawn from the ultrafast pump-probe laser spectroscopy
and spectral hole burning experiments which are
well advanced techniques used for nanostructures and comprise nowadays a standard tool
to determine fast carrier dynamics and spectral and spatial diffusion of the
carriers. A wide-ranging experimental information resulting from the latter
\cite{JONSCHER,SCHLICHTER}  has led to the concept that the
classical phenomenology of relaxation processes breaks down in complex materials.
It has been found that the Debye behavior (\ref{debye})
is hardly ever found in nature and that for many dielectrics the deviations 
from it may be relatively large 
\cite{JONSCHER,HAV,COLE}.
For a long time a major effort has been diverted to a
purely qualitative representation of the shape of the non-Debye dielectric
functions 
in terms of certain mathematical expressions without, in any
way, going into a physical significance of these representations. 
As pointed out by experimental studies almost all dielectric data 
are characterized well enough 
by a few empirical functions
\cite{HAASE,JONSCHER,HAV,DAV,COLE}.
The most popular analytical expression applied to the complex susceptibility
or permittivity data is given by the Havriliak-Negami function 
(\ref{havriliak}).
For $\alpha\!=\!1$ and
$\gamma\!<\!1$, formula (\ref{havriliak}) takes the form (\ref{perm}) of the Cole-Davidson  
function; for $\gamma\!=\!1$ and $\alpha\!<\!1$ it 
takes the form (\ref{cole}) 
of the Cole-Cole  function, and for $\alpha\!=\!1$ and 
$\gamma\!=\!1$ one obtains the classical Debye form (\ref{debye}).
Let us note that time-domain  relaxation function $\phi (t)$ corresponding to formula (\ref{havriliak})	
has the
following series representation 

\begin{equation}
\label{10}
\phi (t)=1- \sum_{n=0}^{\infty}
\frac{(-1)^{n}\,\Gamma(\gamma+n)}{ \Gamma(\gamma)\, n! \,
\Gamma(1+\alpha(\gamma+n))}
\left (\omega_{p}t\right )^{\alpha(\gamma+n)}.
\end{equation}
referred to the generalized Mittag-Leffler distribution. 
In case of the CC function the series representation 
(\ref{10}) is simplified to 
\ben
\phi(t)=1- \sum\limits_{n=0}^{\infty}
\frac{(-1)^{n}}{ \Gamma(1+\alpha(1+n))}
\left (\omega_{p}t\right )^{\alpha(1+n)}
\een
corresponding to the Mittag-Leffler distribution.
The CD relaxation function is referred to the tail function of the
gamma distribution with the scale parameter $\omega_{p}$ and the shape
parameter $\gamma$, given by the density function
\ben
\label{10a}
g_{\gamma}(t) = 
\begin{cases}
\frac{\omega_{p}}{ \Gamma(\gamma)}
(\omega_{p} t)^{\gamma-1} e^{-\omega_{p} t}& \text{for}\;\;\;
t>0,\\	  
0&\text{otherwise}.
\end{cases}
\een
In order to derive  relevant relaxation functions resulting from cases
(\ref{4}) - (\ref{7}), considered above, we use the following
relations:
For any $0<\alpha\leq 1$ 
\ben
\label{11}
\langle e^{-k{\cal S}_{\alpha}}\rangle=
\langle e^{-k{\cal S}_{\alpha}^{'}}\rangle=
e^{-k^{\alpha}}
\een
and for any $0<\gamma\leq 1$ we have \cite{ajurl2003}
\ben
\label{12}
\left\langle e^{-k{\cal B}_{\gamma}}\right\rangle=
\Pr(G_{\gamma}\geq {k})
\een
where ${\cal S}_{1}={\cal S}_{1}^{'}={\cal B}_{1}=1$; 
$G_{1}=E$ is exponentially distributed with mean 1; 
and for $\gamma<1$ the random 
variable $G_{\gamma}$ is distributed according to the
gamma distribution with the shape parameter $\gamma$ and the scale
parameter 1.
\renewcommand{\arraystretch}{1.3}
\doublerulesep 1pt
\tabcolsep3pt
{\small      
\begin{table}[th]
\begin{center}
{
\parbox{8cm}{\centering
{\normalsize\sc Table 1.} 
{\normalsize 
Special cases of the frequency-domain responses (\ref{response}),\\ 
($c_{0}$ is a positive constant.)}
\vskip 0.5truecm}
\begin{tabular}{||c|c|c||c||}
\hline\hline
\multicolumn{3}{||c||}{\sc Assumptions}&
{\sc Response}\\
\cline{1-3}
$\Delta R_{jk}$ & $\Delta T_{jk}$ &  $M_{j}$ & $\phi^{*}(\omega)$\\
\hline\hline
heavy tail & heavy tail & heavy tail & Havriliak-Negami\\
$r\!=\!\alpha$  & $r\!=\!\alpha$ & $r\!=\!\gamma$ & $\langle 
e^{-i(\omega/\omega_{p})  G_{\gamma}^{1/\alpha}{\cal S}_{\alpha}}\rangle$ \\
$c\!=\!\omega_{p}c_{0}/{ k}$ & $c\!=\!c_{0}$ & $c\!>\!0$ & $\alpha, \gamma\!<\! 1$\\
\hline
heavy tail & heavy tail & & Cole-Cole \\
$r\!=\!\alpha$ &  $r\!=\!\alpha$ & $\langle M_{j}\rangle\!<\!\infty$  &
$\langle 
e^{-i(\omega/\omega_{p}) E^{1/\alpha}{\cal S}_{\alpha}}\rangle$\\  
$c\!=\!\omega_{p}c_{0}/{ k}$&
$c\!=\!c_{0}$ & &  $\alpha\!<\! 1$, $\gamma\!=\! 1$\\
\hline
$\langle \Delta R_{jk}\rangle=$ & $\langle \Delta T_{jk}\rangle=$ & heavy tail &
Cole-Davidson\\
$\omega_{p}c_{0}/{ k}<\!\infty$&
$c_{0}\!<\!\infty$&
$r\!=\!\gamma$ & $\langle 
e^{-i(\omega/\omega_{p}) G_{\gamma}}\rangle$\\
& &  $c\!>\!0$& $\alpha\!=\! 1$, $\gamma\!<\! 1$\\
\hline
$\langle \Delta R_{jk}\rangle =$&$\langle \Delta T_{jk}\rangle=$ & & Debye \\
$\omega_{p}c_{0}/{ k}<\!\infty$&
$c_{0}\!<\!\infty$&
$\langle M_{j}\rangle\!<\!\infty$&
$\langle e^{-i(\omega/\omega_{p}) E}\rangle $\\
& & & $\alpha\!=\! 1$, $\gamma\!=\! 1$\\
\hline\hline
\end{tabular}
}
\end{center}
\end{table}
}
Assuming that for any $0<\alpha,\gamma\leq 1$ the 
random variable $G_{\gamma}$ is independent of 
${\cal S}_{\alpha}$, and by using the 
conditional-expectation tools, from formulae (\ref{11}) and (\ref{12})
one obtains
\ben
\left \langle e^{-{ k}
\left (\T{{\cal S}_{\alpha}^{'}\over {\cal S}_{\alpha}}
\left (\T{1\over {\cal B}_{\gamma}}\right )^{1/\alpha}\right )}\right
\rangle=\Pr(G_{\gamma}^{1/\alpha}{\cal S}_{\alpha}\geq { k})
\een
and hence for ${\widetilde R}(t)$ of the form (\ref{4}) -
(\ref{7}) the corresponding relaxation function $\phi (t)$ equals
\ben
\phi (t)=\langle e^{-{ k}{\widetilde R}(t)} \rangle =
\left\langle e^{-t({ k}/A)\left (\T{{\cal S}_{\alpha}^{'}\over {\cal S}_{\alpha}}
\left (\T{1\over {\cal B}_{\gamma}}\right )^{1/\alpha}\right )} \right\rangle \nonumber\\
=\Pr((A/{ k}) G_{\gamma}^{1/\alpha}{\cal S}_{\alpha}\geq t).
\een
where ${ k}$ is an appropriate positive constant. 
For such a relaxation function we have the frequency domain response
of the general form 
\begin{equation}
\label{response}
\phi^{*}(\omega)=\langle 
e^{-i(\omega/\omega_{p}) G_{\gamma}^{1/\alpha}{\cal S}_{\alpha}}\rangle, \;\;\;
\omega_{p}={ k}/A,
\end{equation}
which includes the Havriliak-Negami with its special cases (see Table 1).

\section{Conclusions}
\noindent
Relaxation processes deviating from the usual exponential behavior in time domain 
(and a classical Debye form in the frequency domain) occur in many physical, chemical and
biological systems, such as supercooled liquids, viscoelastic solids, polymer melts and
porous media, membranes and liquid crystals. They are usually described by 
some mathematical functions
  related to the fractional-order differential equations.
  Properties of relaxation 
processes in complex media are also important topics in the long-range transfer of an electron in polymeric and various biological
materials \cite{METZLER,PLONKA,PETROV,MY}. The kinetics of the charge transport in those media is  determined by the nature 
of electronic
coupling  which  for the long distances is mediated by sequential overlaps of
atomic orbitals  of the donor, the intervening medium (bridge), and  the
orbitals of the acceptor.  Internal random motion of the medium may result
in fluctuations of the tunneling barriers between subsequent transfer states
and modulate the electronic coupling. Such an effect
is formally due to the dependence of the electronic coupling on the nuclear
coordinates of the medium and may express possible deviation
from the usual Condon approximation \cite{NEWTON},\cite{STUCH}. The influence
of factors arising
from the static and dynamic fluctuations in electronic coupling have been
discussed in a number of papers (\cite{STUCH},\cite{PANDE},\cite{US} and
references therein) related to
the electron transport in proteins and polymers. All of them have claimed
existence of nontrivial effective coupling resulting from averaging over
environmental disorder. \\
In this paper we have extended previously  proposed model \cite{MY}
 which accounts for fluctuations in long
distance charge transfer hopping. Our approach is based on CTRW formalism
 in the
representation of random sums of {\it {\it i.i.d}}  elements which are deviations from
equilibrium of the atomic coordinates of the intervening medium. In contrast to
other works \cite{METZLER,KLAFTER,SOKOLOV}, the present approach is based on
renewal theory and the limit theorems for random sums
\cite{ZOLOTAREV,meerschaert,WERON,WERON,ajurl2003} of
jumps instead of Tauberian theorems for the two-dimensional Laplace-Fourier
transform. 
Dispersive kinetics, as discussed here, appears in situations when the
environmental fluctuations become comparable or slower in the time-scale of the
overall transition from the reactants' to products' states. 
In  ordinary solvents, the medium relaxation is complete in several picoseconds.
In contrast, in highly viscous liquids the relaxation modes can determine the
rate of the fast ET kinetics. Theoretical treatment of such processes is based
on analysis of slow, reorientational motions of the system which influence 
survival of populations in given electronic states. Our analysis maps the
complex medium reorientation kinetics on a relaxation process of measurable
reactants' and products'  populations for states before and after the charge
transfer. Diffusion of a charge  (and response kinetics of the local
polarization energy) can be then investigated by use of the generalized
relaxation functions defined as moment generating functions for coupled memory random
sums. A non-exponential, non-Debye decay of relaxing modes is a basic feature
arising in such an approach and can be traced to the probabilistic pattern of
contributing lengths of steps and waiting times in the charge hopping process.



\end{document}